\documentclass[11pt]{article}
\usepackage{sectsty}

\usepackage{graphicx}

\usepackage[colorlinks=true]{hyperref}
\hypersetup{urlcolor=blue, citecolor=red, linkcolor=blue}
\usepackage{amsmath,amssymb}
\usepackage{bm}
\usepackage{enumerate}
\usepackage{amsthm}
\usepackage[all,2cell]{xy}
\usepackage{mathrsfs}
\usepackage{mathtools}
\usepackage{tikz-cd}
\usepackage{physics}

\usetikzlibrary{cd}

\usepackage[a4paper, left=25mm, right=25mm, top=25mm, bottom=25mm]{geometry}

\usepackage{marginnote}

\usepackage{imakeidx}
\makeindex[intoc]

\usepackage[nottoc]{tocbibind}

\newtheorem{thm}{Theorem}[section]
\newtheorem{dfn}[thm]{Definition}
\newtheorem{prop}[thm]{Proposition}

\newtheorem{exmpl}[thm]{Example}
\newtheorem{rmrk}[thm]{Remark}

\newcommand\restr[2]{{
	\left.\kern-\nulldelimiterspace 
	#1 
	\right|_{#2} 
}}

\newcommand*{\transp}[2][-3mu]{\ensuremath{\mskip1mu\prescript{\smash{\mathrm t\mkern#1}}{}{\mathstrut#2}}}%

\newcommand{\R}{\mathbb{R}}
\renewcommand{\d}{\mathrm{d}}

\newcommand{\Cinfty}{\mathscr{C}^\infty}
\newcommand{\T}{\mathrm{T}}
\newcommand{\cT}{\mathrm{T}^\ast}

\newcommand{\Lie}{\mathscr{L}}
\newcommand{\X}{\mathfrak{X}}
\newcommand{\vf}{\mathfrak{X}}

\renewcommand{\L}{\mathcal{L}}
\newcommand{\F}{\mathcal{F}}
\renewcommand{\H}{\mathcal{H}}

\newcommand{\W}{\mathcal{W}}
\newcommand{\Reeb}{\mathcal{R}}

\newcommand{\C}{\mathcal{C}}

\renewcommand{\W}{\mathcal{W}}

\newcommand{\parder}[2]{\frac{\partial #1}{\partial #2}}
\newcommand{\dparder}[2]{\dfrac{\partial #1}{\partial #2}}
\newcommand{\tparder}[2]{\partial #1/\partial #2}
\newcommand{\parderr}[3]{\frac{\partial^2 #1}{\partial #2\partial #3}}

\newcommand{\tparderr}[3]{\partial^2 #1/\partial #2\partial #3}

\let\graph\relax
\DeclareMathOperator{\graph}{graph}

\DeclareMathAlphabet{\mathpzc}{OT1}{pzc}{m}{it}
\def\d{\mathrm{d}}

\makeatletter
\let\@fnsymbol\@arabic
\makeatother

\title{Lagrangian--Hamiltonian formalism for time-dependent dissipative mechanical systems}

\author{$^a$Xavier Rivas\footnote{{\bf e}-{\it mail}:
	xavier.rivas@unir.net \quad ORCID: 0000-0002-4175-5157}
	\ and Daniel Torres\footnote{{\bf e}-{\it mail}:
	daniel.torres.moral@gmail.com \quad ORCID: 0000-0001-8355-3783}
	\\[1ex]
\normalsize\itshape\sffamily 
$^a$Escuela Superior de Ingenier\'{\i}a y Tecnolog\'{\i}a,\\
\normalsize\itshape\sffamily 
Universidad Internacional de La Rioja,
Logro\~no, Spain.
}
   
\date{\today}

\begin{document}

\maketitle

\begin{abstract}\noindent
	In this paper we present a unified Lagrangian--Hamiltonian geometric formalism to describe time-dependent contact mechanical systems, based on the one first introduced by K. Kamimura and later formalized by R. Skinner and R. Rusk. This formalism is especially interesting when dealing with systems described by singular Lagrangians, since the second-order condition is recovered from the constraint algorithm. In order to illustrate this formulation, some relevant examples are described in full detail: the Duffing equation, an ascending particle with time-dependent mass and quadratic drag, and a charged particle in a stationary electric field with a time-dependent constraint.
\end{abstract}

\vspace{5mm}

\vspace{5mm}
\noindent\textbf{MSC\,2020 codes:}

37J55; 
70H03, 
70H05, 
53D10, 
70G45, 
53Z05, 
70H45 

\vspace{10mm}

{
\def\baselinestretch{1}
\small
\def\addvspace#1{\vskip 1pt}
\parskip 0pt plus 0.1mm
\tableofcontents
}
\newpage

\section{Introduction}

The Skinner--Rusk formalism was introduced by R. Skinner and R. Rusk in 1983 \cite{Ski1983} (although a previous description in local coordinates had been developed by K. Kamimura in \cite{Kam1982}) in order to deal with mechanical systems described by singular Lagrangian functions. This formulation combines both the Lagrangian and Hamiltonian formalism and this is why it is sometimes called {\it unified } formalism. The Skinner--Rusk formalism has been extended to time-dependent systems \cite{Bar2008,Can2002,Gra2005}, nonholonomic and vakonomic mechanics \cite{Cor2002}, higher-order mechanical systems \cite{Gra1991b,Gra1992,Pri2011,Pri2012}, control systems \cite{Bar2007, Col2010} and field theory \cite{Cam2009,DeLeo2003,Ech2004,Rey2005,Rey2012,Vit2010}. Recently, the Skinner--Rusk unified formalism was extended to contact \cite{DeLeo2020} and $k$-contact \cite{Gra2021} systems. The Skinner--Rusk unified formalism has several advantages. In first place, we recover the second-order condition even if the Lagrangian of the system is singular. We also recover the definition of the Legendre map from the constraint algorithm. Also, both the Lagrangian and Hamiltonian formulations can be recovered from the Skinner--Rusk formalism by projecting onto their respective phase spaces.

The use of contact geometry \cite{Ban2016,Gei2008,Kho2013} to model geometrically the time-dependence in mechanical systems is very well-known \cite{Abr1978,DeLeo2017,Lib1987} and it is, alongside with cosymplectic geometry \cite{CNY-2013}, the natural way to do it. However, in the last decade, the application of contact geometry to the study of dynamical systems has grown significantly \cite{Bra2017a,DeLeo2019b}. This is due to the fact that one can use contact structures to describe many different types of dynamical systems which can not be described by means of symplectic geometry and standard Hamiltonian dynamics in a natural way. The dynamical systems which can be modelled using contact structures include mechanical systems with certain types of damping \cite{Gas2019,LIU2018,Vi-2018}, some systems in quantum mechanics \cite{Cia2018}, circuit theory \cite{Got2016}, control theory \cite{Ram2017} and thermodynamics \cite{Bra2018,Sim2020}, among many others \cite{Bra2020,DeLeo2021c,DeLeo2021b,DeLeo2021d,DeLeo2017,Ese2021,Kho2013,Sus1999}.

Although contact geometry is a suitable framework when working with systems of ordinary differential equations, a generalisation was required in order to deal with systems of partial differential equations describing classical field theories. This generalisation is the so-called $k$-contact structure \cite{Gas2020,Gas2021,Gra2021}. This formulation allows to describe many types of field theories both in the Lagrangian and in the Hamiltonian formalisms. The $k$-contact framework allows us to describe geometrically field theories with damping, some equations from circuit theory, such as the so-called telegrapher's equation, or the Burgers' equation. Recently, a geometric framework has been developed \cite{DeLeo2022} in order to deal with time-dependent mechanical systems with dissipation using the so-called cocontact geometry. It is still an open problem to find a geometric setting to describe non-autonomous field theories with damping.

The main goal of this paper is to extend the Skinner--Rusk formalism to time-dependent contact systems, studying the dynamical equations and the submanifold where they are consistent, and showing that the Lagrangian and Hamiltonian formalisms can be recovered from this mixed formalism. In first place, we introduce the phase space of this formulation: the Pontryagin bundle $\W = \R\times\T Q\times\cT Q\times\R$. This manifold has a natural precocontact structure $(\d t,\eta)$ inherited from the natural cocontact structure of $\R\times\cT Q\times\R$ (see \cite{DeLeo2022}). The Hamiltonian energy associated to a Lagrangian function $L\in\Cinfty(\R\times\T Q\times\R)$ is defined as
$$ \H = p_iv^i - L(t,q^i,v^i,s)\,. $$
Since the Hamiltonian system $(\W,\d t,\eta,\H)$ is singular, we need to implement a constraint algorithm in order to find a submanifold where the Hamiltonian equations are consistent. In the first iteration of the constraint algorithm we recover the second-order condition, even if the Lagrangian function is singular (in the Lagrangian formalism, we only recover the {\sc sode} condition if the Lagrangian is regular). The first constraint submanifold is the graph of the Legendre map $\F L$. If the Lagrangian function is regular the constraint algorithm ends in one step and we obtain the usual results by projecting the dynamics onto the Lagrangian and Hamiltonian phase spaces. If the Lagrangian function is singular, the constraint algorithm is related to the usual Lagrangian and Hamiltonian constraint algorithms (imposing the {\sc sode} condition in the Lagrangian case).

The structure of the present paper is as follows. In Section 2, we review the basics on cocontact geometry, which is an extension of both contact and cosymplectic geometry. This geometric framework allows us to develop a Hamiltonian and a Lagrangian formulation for time-dependent contact systems \cite{DeLeo2022}. Section 3 is devoted to present the Skinner--Rusk unified formulation for cocontact systems. We begin by introducing the Pontryagin bundle and its natural precocontact structure and state the Lagrangian--Hamiltonian problem. In Section 4 we recover both the Lagrangian and Hamiltonian formalisms and see that they are equivalent to the Skinner--Rusk formalisms (imposing the second order-condition if the Lagrangian is singular). Finally, in Section 5 some examples are studied in full detail. These examples are the Duffing equation \cite{Guc1983,Wig2003}, an ascending particle with time-dependent mass and quadratic drag, and a charged particle in a stationary electric field with a time-dependent constraint.

Throughout this paper, all the manifolds are real, second countable and of class $\Cinfty$. Mappings are assumed to be smooth and the sum over crossed repeated indices is understood.

\section{Review on time-dependent contact systems}

In this first section we will briefly review the basics on cocontact manifolds introduced in \cite{DeLeo2022} and how this structure can be used to geometrically describe time-dependent contact mechanical systems.

\subsection{Cocontact geometry}

\begin{dfn}
    A \textbf{cocontact structure} on a $(2n+2)$-dimensional manifold $M$ is a couple of 1-forms $(\tau, \eta)$ on $M$ such that $\d\tau = 0$ and $\tau\wedge\eta\wedge(\d{\eta})^n\neq 0$.
    In this case, $(M, \tau, \eta)$ is said to be a \textbf{cocontact manifold}.
\end{dfn}

\begin{exmpl}\label{ex:canonical-cocontact-manifold}\rm
    Let $Q$ be an $n$-dimensional smooth manifold with local coordinates $(q^i)$. Let $(q^i, p_i)$ be the induced natural coordinates on its cotangent bundle $\cT Q$.
    Consider the product manifolds $\R\times\cT Q$, $\cT Q\times\R$ and $\R\times\cT Q\times\R$ with natural coordinates $(t, q^i, p_i)$,$(q^i, p_i, s)$ and $(t, q^i, p_i, s)$ respectively. Let us also define the following projections:

    \begin{center}
        \begin{tikzcd}
            & \R\times\cT Q\times\R \arrow[dl, swap, "\rho_1"] \arrow[dr, "\rho_2"] \arrow[dd, "\pi"] & \\
            \R\times\cT Q \arrow[dr, swap, "\pi_2"] & & \cT Q\times\R \arrow[dl, "\pi_1"] \\
            & \cT Q &
        \end{tikzcd}
    \end{center}
 
    Now consider $\theta_0\in\Omega^1(\cT Q)$ be the canonical 1-form of the cotangent bundle with local expression $\theta_0=p_i\d q^i$ and let $\theta_1 = \pi_1^*\theta_0$ and $\theta_2 = \pi_2^*\theta_0$.
     
    Then we have that $(\d t, \theta_2)$ is a cosymplectic structure in $\R\times\cT Q$ and $\eta_1 = \d s - \theta_1$ is a contact form on $\cT Q\times\R$. Furthermore, considering the one-forms in $\R\times\cT Q\times\R$ given by $\theta = \rho_1^*\theta_2 = \rho_2^*\theta_1 = \pi^*\theta_0$, $\tau = \d t$ and $\eta = \d s - \theta$, we have that $(\tau, \eta)$ is a cocontact structure in $\R\times\cT Q\times\R$ with local expression:
    $$ \tau = \d t\,,\qquad \eta = \d s - p_i\d q^i\,. $$
\end{exmpl}

\bigskip
In a cocontact manifold $(M, \tau, \eta)$ we have the so called \textbf{flat isomorphism}
$$
    \begin{array}{rccl}
        \flat\colon & \T M & \longrightarrow & \cT M \\
        & v & \longmapsto & (i(v)\tau)\tau + i(v)\d\eta + \left(i(v)\eta\right)\eta\,,
    \end{array}
$$
which can be extended to a morphism of $\Cinfty(M)$-modules:
$$ \flat:X\in\X(M)\longmapsto (i(X)\tau)\tau + i(X)\d\eta + \left(i(X)\eta\right)\eta\in\Omega^1(M)\,. $$

\begin{prop}\label{prop:Reeb-vector-fields}
    Given a cocontact manifold $(M, \tau, \eta)$ there exist two vector fields $R_t$, $R_s$ on $M$ such that
    $$
        \begin{cases}
            i(R_t)\tau = 1\,,\\
            i(R_t)\eta = 0\,,\\
            i(R_t)\d\eta = 0\,,
        \end{cases}
        \qquad
        \begin{cases}
            i(R_s)\tau = 0\,,\\
            i(R_s)\eta = 1\,,\\
            i(R_s)\d\eta = 0\,.
        \end{cases}
    $$
    Equivalently, they can be defined as $R_t = \flat^{-1}(\tau)$ and $R_s = \flat^{-1}(\eta)$. The vector fields $R_t$ and $R_s$ are called \textbf{time and contact Reeb vector fields} respectively.
\end{prop}

Moreover, on a cocontact manifold we also have the \textbf{canonical} or \textbf{Darboux} coordinates, as the following theorem establishes:

\begin{thm}[Darboux theorem for cocontact manifolds]
    Given a cocontact manifold $(M, \tau, \eta)$, for every $p\in M$ exists a local chart $(U; t, q^i, p_i, s)$ containing $p$ such that
    $$ \restr{\tau}{U} = \d t\,, \qquad \restr{\eta}{U} = \d s - p_i\d q^i\,.$$
    In Darboux coordinates, the Reeb vector fields read $R_t = \tparder{}{t}$, $R_s = \tparder{}{s}$.
\end{thm}

\subsection{Cocontact Hamiltonian systems}

\begin{dfn}
	A \textbf{cocontact Hamiltonian system} is family $(M,\tau,\eta,H)$ where $(\tau,\eta)$ is a cocontact structure on $M$ and $H:M\to\R$ is a Hamiltonian function. The \textbf{cocontact Hamilton equations} for a curve $\psi\colon I\subset \R\to M$ are
	\begin{equation}\label{eq:Ham-eq-cocontact-sections}
		\begin{dcases}
			i(\psi')\d \eta = \d H-(\Lie_{R_s}H)\eta-(\Lie_{R_t}H)\tau\,,
			\\
			i(\psi')\eta = -H\,,
			\\
			i(\psi')\tau = 1\,,
		\end{dcases}
	\end{equation}
	where $\psi'\colon I\subset\R\to\T M$ is the canonical lift of $\psi$ to the tangent bundle $\T M$. The \textbf{cocontact Hamiltonian equations} for a vector field $X\in\X(M)$ are:
	\begin{equation}\label{eq:Ham-eq-cocontact-vectorfields}
		\begin{dcases}
			   i(X)\d \eta = \d H-(\Lie_{R_s}H)\eta-(\Lie_{R_t}H)\tau\,,
			\\
			i(X)\eta = -H\,,
			\\
			i(X)\tau = 1\,,
		\end{dcases}
	\end{equation}
	or equivalently, $\flat(X)=\d H-\left(\Lie_{R_s}H+H\right)\eta+\left(1-\Lie_{R_t}H\right)\tau$. The unique solution to this equations is called the \textbf{cocontact Hamiltonian vector field}.
\end{dfn}

Given a curve $\psi$ with local expression $\psi(r)=(f(r),q^i(r),p_i(r),s(r))$, the third equation in \eqref{eq:Ham-eq-cocontact-sections} imposes that $f(r)=r + cnt$, thus we will denote $r\equiv t$, while the other equations read:
\begin{equation}\label{eq:Hamilton-cocontact}
   \begin{dcases}
		\dot q^i =\frac{\partial H}{\partial p_i}\,,
		\\
	   \dot p_i =-\left(\frac{\partial H}{\partial q^i}+p_i\frac{\partial H}{\partial s}\right)\,,
		\\
	   \dot s =p_ i\frac{\partial H}{\partial p_i}-H\,.		
	\end{dcases}	  
\end{equation}
On the other hand, the local expression of the cocontact Hamiltonian vector field is
$$ X = \parder{}{t} + \parder{H}{p_i}\parder{}{q^i} - \left(\parder{H}{q^i} + p_i\parder{H}{s}\right)\parder{}{p_i} + \left(p_i\parder{H}{p_i} - H\right)\parder{}{s}\,. $$

\subsection{Cocontact Lagrangian systems}\label{sec:Lagrangian-formalism}

Given a smooth $n$-dimensional manifold $Q$, consider the product manifold $\R\times\T Q\times \R$ equipped with canonical coordinates $(t, q^i, v^i, s)$. We have the canonical projections
\begin{align*}
	\tau_1&\colon \R\times\T Q\times\R\to\R\ ,&& \tau_1(t, v_q, s) = t\,,\\
	\tau_2&\colon\R\times\T Q\times\R\to\T Q\ ,&& \tau_2(t, v_q, s) = v_q\,,\\
	\tau_3&\colon\R\times\T Q\times\R\to\R\ ,&& \tau_3(t, v_q, s) = s\,,\\
	\tau_0&\colon\R\times\T Q\times\R\to \R\times Q\times\R\ ,&& \tau_0(t, v_q, s) = (t, q, s)\,. 
\end{align*}

The usual geometric structures of the tangent bundle can be naturally extended to the cocontact Lagrangian phase space $\R\times\T Q\times\R$. In particular, the vertical endomorphism of $\T(\T Q)$ yields a \textbf{vertical endomorphism} $\mathcal{J}\colon\T(\R\times\T Q\times\R)\to\T(\R\times\T Q\times\R)$. In the same way, the Liouville vector field on the fibre bundle $\T Q$ gives a \textbf{Liouville vector field} $\Delta\in\X(\R\times\T Q\times\R)$. The local expressions of these objects in Darboux coordinates are
$$ {\cal J} = \frac{\partial}{\partial v^i} \otimes \d q^i \,,\quad \Delta =  v^i\, \frac{\partial}{\partial v^i} \,. $$

\begin{dfn}\label{dfn:holonomic-path}
    Given a path ${\bf c} \colon\R \rightarrow \R\times Q\times\R$ with ${\bf c} = (\mathbf{c}_1,\mathbf{c}_2,\mathbf{c}_3)$, the \textbf{prolongation} of ${\bf c}$ to $\R\times\T Q\times\R$ is the path ${\bf \tilde c} =  (\mathbf{c}_1,\mathbf{c}_2',\mathbf{c}_3) \colon \R \longrightarrow \R\times\T Q \times \R$, where $\mathbf{c}_2'$ is the velocity of~$\mathbf{c}_2$. Every path ${\bf \tilde c}$ which is the prolongation of a path ${\bf c} \colon\R \rightarrow \R\times Q\times\R$ is called \textbf{holonomic}. A vector field 
    $\Gamma \in \X(\R\times\T Q \times \R)$ satisfies the \textbf{second-order condition} (it is a {\sc sode}) if all of its integral curves are holonomic.
\end{dfn}

The vector fields satisfying the second-order condition can be characterized by means of the canonical structures $\Delta$ and $\mathcal{J}$ introduced above, since $X$ is a {\sc sode} if and only if ${\cal J}( \Gamma) = \Delta$.

Taking canonical coordinates, if ${\bf c}(r)=(t(r), c^i(r), s(r))$, its prolongation to $\R\times\T Q\times\R$ is
$$ {\bf \tilde c}(r) = \left(t(r), c^i(r),\frac{d c^i}{d r}(r), s(r) \right) \,. $$
The local expression of a {\sc sode} in natural coordinates $(t,q^i,v^i,s)$ is
\begin{equation}\label{eq:sode-local-expression}
    \Gamma = f\parder{}{t} + v^i \parder{}{q^i} + G^i \parder{}{v^i} + g\,\frac{\partial}{\partial s}\,.
\end{equation}
Thus, a {\sc sode} defines a system of
differential equations of the form
$$
    \frac{\d t}{\d r} = f(t, q, \dot q, s)\,,\quad \frac{\d^2 q^i}{\d r^2} = G^i(t, q,\dot q,s) \,, \quad \frac{\d s}{\d r} = g(t, q,\dot q,s)  \:.
$$

\begin{dfn}\label{dfn:lagrangian}
    A \textbf{Lagrangian function} is a function $\L\in\Cinfty(\R\times\T Q\times\R)$. The \textbf{Lagrangian energy} associated to $\L$ is the function $E_\L = \Delta(\L)-\L$. The \textbf{Cartan forms} associated to $\L$ are
	\begin{equation}\label{eq:thetaL}
		\theta_\L = \transp{\cal J}\circ\d\L\in \Omega^1(\R\times\T Q\times\R)\,,\quad\omega_\L = -\d\theta_\L\in \Omega^2(\R\times\T Q\times\R) \,.
	\end{equation}
	The \textbf{contact Lagrangian form} is
	$$ \eta_\L=\d s-\theta_\L\in\Omega^1(\R\times\T Q\times\R)\,. $$
	Notice that $\d\eta_\L=\omega_\L$. The couple $(\R\times\T Q\times\R,\L)$ is a \textbf{cocontact Lagrangian system}.
\end{dfn}

The local expressions of these objects are
\begin{gather*}
	E_\L = v^i\parder{\L}{v^i} - \L\,,\qquad \eta_\L = \d s - \frac{\partial\L}{\partial v^i} \,\d q^i \,,\\
	\d\eta_\L = -\parderr{\L}{t}{v^i}\d t\wedge\d q^i -\parderr{\L}{q^j}{v^i}\d q^j\wedge\d q^i -\parderr{\L}{v^j}{v^i}\d v^j\wedge\d q^i -\parderr{\L}{s}{v^i}\d s\wedge\d q^i \,.
\end{gather*}

Not all cocontact Lagrangian systems $(\R\times\T Q\times\R,\L)$ result in the family $(\R\times\T Q\times\R,\tau=\d t,\eta_\L, E_\L)$ being a cocontact Hamiltonian system because the condition $\tau\wedge\eta\wedge(\d\eta_\L)^n\neq 0$ is not always fulfilled. The Legendre map characterizes the Lagrangian functions will result in cocontact Hamiltonian systems.

\begin{dfn}\label{dfn:Legendre-map}
    Given a Lagrangian function $\L \in\Cinfty(\R\times \T Q\times\R)$, the \textbf{Legendre map} associated to $\L$ is its fibre derivative, considered as a function on the vector bundle $\tau_0 \colon\R\times \T Q\times\R \to \R\times Q \times \R$; that is, the map $\F\L \colon \R\times\T Q \times\R \to \R\times\cT Q \times \R$ with local expression
    $$
        \F\L (t,v_q,s) = \left( t,\F\L(t,\cdot,s) (v_q),s \right)\,,
    $$
    where $\F\L(t,\cdot,s)$ is the usual Legendre map associated to the Lagrangian $\L(t,\cdot,s)\colon\T Q\to\R$ with $t$ and $s$ freezed.
\end{dfn}

The Cartan forms can also be defined as $\theta_\L={\cal FL}^{\;*}(\pi^*\theta_0)$ and $\omega_\L={\cal FL}^{\;*}(\pi^*\omega_0)$, where $\theta_0$ and $\omega_0 = -\d\theta_0$ are the canonical one- and two-forms of the cotangent bundle and $\pi$ is the natural projection $\pi\colon\R\times\cT Q\times\R\to\cT Q$ (see Example \ref{ex:canonical-cocontact-manifold}).

\begin{prop}\label{prop:regular-Lagrangian}
    Given a Lagrangian function $\L$ the following statements are equivalent:
    \begin{enumerate}[{\rm(i)}]
        \item The Legendre map $\F \L$ is a local diffeomorphism.
        \item The fibre Hessian $\F^2\L \colon\R\times\T Q\times\R \longrightarrow (\R\times\cT Q\times\R)\otimes (\R\times\cT Q\times\R)$ of $\L$ is everywhere nondegenerate (the tensor product is understood to be of vector bundles over $\R\times Q \times \R$).
        \item The family $(\R\times\T Q\times\R,\d t, \eta_\L)$ is a cocontact manifold.
    \end{enumerate}
\end{prop}
This can be checked using that $\F\L(t,q^i,v^i, s)=\left(t,q^i,\tparder{\L}{v^i}, s\right)$ and $\F^2 \L(t,q^i,v^i,s) = (t,q^i,W_{ij},s)$, where $W_{ij} = \left( \tparderr{\L}{v^i}{v^j}\right)$.

A Lagrangian function $\L$ is \textbf{regular} if the equivalent statements in the previous proposition hold. Otherwise $\L$ is \textbf{singular}. Moreover, $\L$ is \textbf{hyperregular} if $\F\L$ is a global diffeomorphism. Thus, every \emph{regular} cocontact Lagrangian system yields the cocontact Hamiltonian system $(\R\times\T Q\times\R,\d t, \eta_\L, E_\L)$.

Given a regular cocontact Lagrangian system $(\R\times\T Q\times\R,\L)$, the \textbf{Reeb vector fields} 
$R_t^\L,R_s^\L\in\X(\R\times\T Q\times\R)$ are uniquely determined by the relations
$$ \begin{dcases}
		i(R_t^\L)\d t = 1\,,\\
		i(R_t^\L)\eta_\L = 0\,,\\
		i(R_t^\L)\d\eta_\L = 0\,,
	\end{dcases}\qquad\begin{dcases}
		i(R_s^\L)\d t = 0\,,\\
		i(R_s^\L)\eta_\L = 1\,,\\
		i(R_s^\L)\d\eta_\L = 0\,,
	\end{dcases} $$
and their local expressions are
\begin{equation*}
    R_t^\L = \parder{}{t} - W^{ij}\parderr{\L}{t}{v^j}\parder{}{v^i}\,,\qquad
    R_s^\L = \parder{}{s} - W^{ij}\parderr{\L}{s}{v^j}\parder{}{v^i}\,,
\end{equation*}
where $W^{ij}$ is the inverse of the Hessian matrix of the Lagrangian $\L$, namely $W^{ij}W_{jk}=\delta^i_k$.

If the Lagrangian $\L$ is singular, the Reeb vector fields are not uniquely determined, actually, they may not even exist \cite{DeLeo2022}.

\subsection{The Herglotz--Euler--Lagrange equations}

\begin{dfn}\label{dfn:Euler-Lagrange-equations}
	Given a regular cocontact Lagrangian system $(\R\times\T Q\times\R,\L)$ the \textbf{Herglotz--Euler--Lagrange equations} for a holonomic curve ${\bf\tilde c}\colon I\subset\R \to\R\times\T Q\times\R$ are
	\begin{equation}\label{eq:Euler-Lagrange-curve}
		\begin{dcases}
			i({\bf\tilde c}')\d\eta_\L = \left(\d E_\L - (\Lie_{R_t^\L}E_\L)\d t - (\Lie_{R_s^\L}E_\L)\eta_\L\right)\circ{\bf\tilde c}\,,\\
			i({\bf\tilde c}')\eta_\L = - E_\L\circ{\bf\tilde c}\,,\\
			i({\bf\tilde c}')\d t = 1\,,
		\end{dcases}
	\end{equation}
	where ${\bf\tilde c}'\colon I\subset\R\to\T(\R\times\T Q\times\R)$ is the canonical lift of ${\bf\tilde c}$ to $\T(\R\times\T Q\times\R)$. The \textbf{cocontact Lagrangian equations} for a vector field $X_\L\in\X(\R\times\T Q\times\R)$ are 
	\begin{equation}\label{eq:Euler-Lagrange-field}
		\begin{cases}
			i(X_\L)\d\eta_\L = \d E_\L-(\Lie_{R_t^\L}E_\L)\d t-(\Lie_{R_s^\L}E_\L)\eta_\L\,,\\
			i(X_\L)\eta_\L = -E_\L \,,\\
			i(X_\L)\d t = 1\,.
		\end{cases}
	\end{equation}
	The only vector field solution to these equations is the \textbf{cocontact Lagrangian vector field}.
\end{dfn}

\begin{rmrk}\rm
    The cocontact Lagrangian vector field of a regular cocontact Lagrangian system $(\R\times\T Q\times\R,\L)$ is the cocontact Hamiltonian vector field of the cocontact Hamiltonian system $(\R\times\T Q\times\R,\d t,\eta_\L,E_\L)$.
\end{rmrk}

Given a holonomic curve
${\bf\tilde c}(r)=(t(r), q^i(r),\dot q^i(r),s(r))$,
equations \eqref{eq:Euler-Lagrange-curve} read
\begin{align}
	\dot t &= 1\,,\label{eq:EL-curve-coords-1}\\
	\dot s &= \L\,,\label{eq:EL-curve-coords-2}\\
	\dot t\parderr{\L}{t}{v^i} + \dot q^j\parderr{\L}{q^j}{v^i} + \ddot q^j\parderr{\L}{v^j}{v^i} + \dot s\parderr{\L}{s}{v^i} - \parder{\L}{q^i} = \frac{\d}{\d r}\left(\parder{\L}{v^i}\right) - \parder{\L}{q^i} &= \parder{\L}{s}\parder{\L}{v^i}\,.\label{eq:EL-curve-coords-3}
\end{align}
The fact that $\dot t = 1$ justifies the usual identification $t\equiv r$. For a vector field $X_\L$ with local expression $X_\L = f\dparder{}{t} + F^i\dparder{}{q^i} + G^i\dparder{}{v^i} + g\dparder{}{s}$, equations \eqref{eq:Euler-Lagrange-field} are
\begin{align}
	(F^j - v^j)\parderr{\L}{t}{v^j} &= 0\,,\label{eq:EL-field-coords-1}\\
	f\parderr{\L}{t}{v^i} + F^j\parderr{\L}{q^j}{v^i} + G^j\parderr{\L}{v^j}{v^i} + g\parderr{\L}{s}{v^i} - \parder{\L}{q^i} - (F^j - v^j)\parderr{\L}{q^i}{v^j} &= \parder{\L}{s}\parder{\L}{v^i}\,,\label{eq:EL-field-coords-2}\\
	(F^j - v^j)\parderr{\L}{v^i}{v^j} &= 0\,,\label{eq:EL-field-coords-3}\\
	(F^j - v^j)\parderr{\L}{s}{v^j} &= 0\,,\label{eq:EL-field-coords-4}\\
	\L + \parder{\L}{v^j}(F^j-v^j) - g &= 0\,,\label{eq:EL-field-coords-5}\\
	f &= 1\label{eq:EL-field-coords-6}\,.
\end{align}

\begin{thm}\label{thm:regular-lagrangian}
    If $\L$ is a regular Lagrangian, $X_\L$ is a {\sc sode} and equations \eqref{eq:EL-field-coords-1}--\eqref{eq:EL-field-coords-6} become
    \begin{equation*}
        f = 1\,,\quad g = \L\,,\quad \parderr{\L}{t}{v^i} + v^j\parderr{\L}{q^j}{v^i} + G^j\parderr{\L}{v^j}{v^i} + \L\parderr{\L}{s}{v^i} - \parder{\L}{q^i}
        = \parder{\L}{s}\parder{\L}{v^i}\,,
    \end{equation*}
    which, for the integral curves of $X_\L$, are the \textbf{Herglotz--Euler--Lagrange equations} \eqref{eq:EL-curve-coords-1}, \eqref{eq:EL-curve-coords-2} and \eqref{eq:EL-curve-coords-3}. This {\sc sode} $X_\L\equiv\Gamma_\L$ is the \textbf{Herglotz--Euler--Lagrange vector field} for the Lagrangian $\L$.
\end{thm}

The coordinate expression of the Herglotz--Euler--Lagrange vector field is
\begin{equation*}
    \Gamma_\L = \parder{}{t} + v^i\parder{}{q^i} + W^{ji}\left( \parder{\L}{q^j} - \parderr{\L}{t}{v^j} - v^k\parderr{\L}{q^k}{v^j} - \L\parderr{\L}{s}{v^j} + \parder{\L}{s}\parder{\L}{v^j} \right)\parder{}{v^i} + \L\parder{}{s}\,.
\end{equation*}
An integral curve of $\Gamma_\L$ fulfills the Herglotz--Euler--Lagrange equation for dissipative systems:
$$ \frac{\d}{\d t}\left(\parder{\L}{v^i}\right) - \parder{\L}{q^i} =\parder{\L}{s}\parder{\L}{v^i}\,,\qquad
\dot s = \L\,.
$$

\section{Skinner--Rusk formalism}

Consider a cocontact Lagrangian system with configuration space $\R\times Q\times\R$, where $Q$ is an $n$-dimensional manifold, equipped with coordinates $(t, q^i, s)$. Consider the product bundles $\R\times\T Q\times\R$ with natural coordinates $(t, q^i, v^i, s)$ and $\R\times\cT Q\times\R$ with natural coordinates $(t, q^i, p_i, s)$, and the natural projections
\begin{gather*}
    \tau_2\colon\R\times\T Q\times\R\to\T Q\ ,\qquad \tau_0\colon\R\times\T Q\times\R\to\R\times Q\times\R\,,\\
    \pi_2\colon\R\times\cT Q\times\R\to\cT Q\ ,\qquad \pi_0\colon\R\times\cT Q\times\R\to\R\times Q\times\R\,.
\end{gather*}

Let $\theta_0\in\Omega^1(\cT Q)$ denote the Liouville 1-form of the cotangent bundle and let $\omega_0 = -\d\theta_0\in\Omega^2(\cT Q)$ be the canonical symplectic form of $\cT Q$. The local expressions of $\theta_0$ and $\omega_0$ are
$$ \theta_0 = p_i\d q^i\ ,\qquad \omega_0 = \d q^i\wedge\d p_i\,.$$
We will denote by $\theta = \pi_2^\ast\theta_0\in\Omega^1(\R\times\cT Q\times\R)$ and $\omega = \pi_2^\ast\omega_0\in\Omega^2(\R\times\cT Q\times\R)$ the pull-backs of $\theta_0$ and $\omega_0$ to $\R\times\cT Q\times\R$.

\begin{dfn}
    The extended Pontryagin bundle is the Whitney sum
    $$ \W = \R\times\T Q\times_Q\cT Q\times\R $$
    equipped with the natural submersions
    \begin{align*}
        &\rho_1\colon\W\to\R\times\T Q\times\R\,,\\
        &\rho_2\colon\W\to\R\times\cT Q\times\R\,,\\
        &\rho_0\colon\W\to\R\times Q\times\R\,.
    \end{align*}
\end{dfn}

The extended Pontryagin bundle is endowed with natural coordinates $(t, q^i, v^i, p_i, s)$.

\begin{dfn}
    A path $\gamma\colon\R\to\W$ is \textbf{holonomic} if the path $\rho_1\circ\gamma\colon\R\to\R\times\T Q\times\R$ is holonomic, i.e., it is the prolongation to $\R\times\T Q\times\R$ of a path $\R\to\R\times Q\times\R$.
    
    A vector field $X\in\X(\W)$ satisfies the \textbf{second-order condition} (or it is a \textsc{sode}) if its integral curves are holonomic in $\W$.
\end{dfn}

A holonomic path in $\W$ has local expression
$$ \gamma(\tau) = \left( t(\tau), q^i(\tau), \dot q^i(\tau), p_i(\tau), s(\tau) \right)\,. $$
The local expression of a \textsc{sode} in $\W$ is
$$ X = f\parder{}{t} + v^i\parder{}{q^i} + F^i\parder{}{v^i} + G_i\parder{}{p_i} + g\parder{}{s}\,.$$

In the extended Pontryagin bundle $\W$ we have the following canonical structures:

\begin{dfn}
    \begin{enumerate}
        \item The \textbf{coupling function} in $\W$ is the map $\C\colon\W\to\R$ given by
        $$ \C(w) = i({\rm v}_q){\rm p}_q\,, $$
        where $w = (t, {\rm v}_q, {\rm p}_q, s)\in\W$, $q\in Q$, ${\rm v}_q\in\T Q$ and ${\rm p}_q\in\cT Q$.
        \item The \textbf{canonical 1-form} is the $\rho_0$-semibasic form $\Theta = \rho_2^\ast\theta \in \Omega^1(\W)$. The \textbf{canonical 2-form} is $\Omega = -\d\Theta = \rho_2^\ast\omega \in \Omega^2(\W)$.
        \item The \textbf{canonical precontact 1-form} is the $\rho_0$-semibasic form $\eta = \d s - \Theta \in \Omega^1(\W)$.
    \end{enumerate}
\end{dfn}
In natural coordinates,
$$ \Theta = p_i\d q^i\ ,\quad \eta = \d s - p_i\d q^i\ ,\quad \d\eta = \d q^i\wedge\d p_i = \Omega\,. $$

\begin{dfn}\label{dfn:Hamiltonian function}
    Let $L\in\Cinfty(\R\times\T Q\times\R)$ be a Lagrangian function and consider $\L = \rho^*_1L\in\Cinfty(\W)$. The \textbf{Hamiltonian function associated to $\L$} is the function
    $$ \H = \C - \L = p_iv^i - \L(t, q^j, v^j, s) \in \Cinfty(\W) $$
\end{dfn}

\begin{rmrk}\rm
Notice that the couple $(\d t,\eta)$ is a precocontact structure in the Pontryagin bundle $\W$. Hence, $(\W, \d t, \eta)$ is a precocontact manifold and $(\W,\d t,\eta,\H)$ is a precocontact Hamiltonian system. These notions were introduced in \cite{DeLeo2022}. Thus, we do not have a unique couple $(\Reeb_t,\Reeb_s)$ of Reeb vector fields. In fact, in natural coordinates, the general solution to \eqref{prop:Reeb-vector-fields} is
\begin{align*}
    \Reeb_t = \parder{}{t} + F^i\parder{}{v^i}\,,\\
    \Reeb_s = \parder{}{s} + G^i\parder{}{v^i}\,,
\end{align*}
where $F^i,G^i\in\Cinfty(\W)$ are arbitrary function. Despite this fact, the formalism is independent on the choice of the Reeb vector fields.
Since the Pontryagin bundle $\W$ is trivial over $\R\times\R$, the vector fields $\tparder{}{t},\tparder{}{s}$ can be canonically lifted to $\W$ and used as Reeb vector fields.
\end{rmrk}

\begin{dfn}
    The \textbf{Lagrangian--Hamiltonian problem} associated to the precocontact Hamiltonian system $(\W, \d t, \eta, \H)$ consists in finding the integral curves of a vector field $Z\in\X(\W)$ such that
    $$ \flat(Z) = \d \H - (\Lie_{\Reeb_s}\H + \H)\eta - (1 - \Lie_{\Reeb_t}\H)\d t\,. $$
    Equivalently,
    \begin{equation}
        \label{eqn:hamiltonian_equations}
        \begin{cases}
        i(Z)\d\eta = \d\H - (\Lie_{\Reeb_s}\H)\eta - (\Lie_{\Reeb_t}\H)\d t\,, \\
        i(Z)\eta = -\H\,,\\
        i(Z)\d t = 1\,.
        \end{cases}
    \end{equation}
    Thus, the integral curves $\gamma\colon I\subset\R\rightarrow\W$ of $Z$ are solutions to the system of equations
    \begin{equation}\label{eqn:hamiltonian_equations_coordinates}
        \begin{cases}
        i(\gamma')\d\eta = (d\H - (\Lie_{\Reeb_s}\H)\eta - (\Lie_{\Reeb_t}\H)\d t)\circ\gamma\,, \\
        i(\gamma')\eta = -\H\circ\gamma\,,\\
        i(\gamma')\d t = 1\,.
        \end{cases}
    \end{equation}
\end{dfn}

Since $(\W, \d t, \eta, \H)$ is a precocontact system, equations \eqref{eqn:hamiltonian_equations} may not be consistent everywhere in $\W$. In order to find a submanifold $\W_f\hookrightarrow\W$ (if possible) where equations \eqref{eqn:hamiltonian_equations} have consistent solutions, a constraint algorithm is needed. The implementation of this algorithm is described below.

Consider the natural coordinates $(t, q^i, v^i, p_i, s)$ in $\W$ and the vector field $Z\in\X(\W)$ with local expression
$$ Z = A\parder{}{t} + B^i\parder{}{q^i} + C^i\parder{}{v^i} + D_i\parder{}{p_i} + E\parder{}{s}\,. $$
The left-hand side of equations \eqref{eqn:hamiltonian_equations} read
\begin{align*}
    i(Z)\d\eta &= B^i\d p_i - D_i\d q^i\,,\\
    i(Z)\eta &= E - p_i B^i\,,\\
    i(Z)\d t &= A\,.
\end{align*}
On the other hand, we have
\begin{align*}
    \d\H &= v^i\d p_i + \left(p_i - \parder{\L}{v^i}\right)\d v^i - \parder{\L}{t}\d t - \parder{\L}{q^i}\d q^i - \parder{\L}{s}\d s\,,\\
    (\Lie_{\Reeb_s}\H)\eta &= -\parder{\L}{s}(\d s - p_i\d q^i)\,,\\
    (\Lie_{\Reeb_t}\H)\d t &= -\parder{\L}{t}\d t\,.
\end{align*}
Thus, the second equation in \eqref{eqn:hamiltonian_equations} gives
\begin{equation}\label{eq:SR-second}
    E = (B^i - v^i)p_i + \L\,,
\end{equation}
the third equation in \eqref{eqn:hamiltonian_equations} reads
\begin{equation}\label{eq:SR-third}
    A = 1\,,    
\end{equation}
and the first equation in \eqref{eqn:hamiltonian_equations} gives the conditions
\begin{align}
	B^i &= v^i & &(\mbox{coefficients in }\d p_i)\,,\label{eq:SR-contact-one}\\
	p_i &= \parder{\L}{v^i} & &(\mbox{coefficients in }\d v^i)\,,\label{eq:SR-contact-two}\\
	D_i &= \parder{\L}{q^i} + p_i\parder{\L}{s} & &(\mbox{coefficients in }\d q^i)\,.\label{eq:SR-contact-three}
\end{align}
Notice that
\begin{itemize}
    \item Equations \eqref{eq:SR-contact-one} are the \textsc{sode} conditions. Hence, the vector field $Z$ is a \textsc{sode}. Then, it is clear that the holonomy condition arises straightforwardly from the Skinner--Rusk formalism.
    \item Conditions \eqref{eq:SR-contact-two} are constraint functions defining the \textbf{first constraint submanifold} $\W_1\hookrightarrow\W$. It is important to notice that the submanifold $\W_1$ is the graph of the Legendre map $\F L$ defined previously \ref{dfn:Legendre-map}:
    $$ \W_1 = \{({\rm v}_q, \F L(t,{\rm v}_q, s))\in\W\mid({t,\rm v}_q,s)\in\R\times\T Q\times\R\}\,. $$
    This implies that the Skinner--Rusk formalism implies the definition of the Legendre map.
\end{itemize}

In virtue of conditions \eqref{eq:SR-contact-one}, \eqref{eq:SR-contact-two}, \eqref{eq:SR-contact-three}, the vector fields $Z$ solution to equations \eqref{eqn:hamiltonian_equations} have the local expression
$$ Z = \parder{}{t} + v^i\parder{}{q^i} + C^i\parder{}{v^i} + \left(\parder{\L}{q^i} + p_i\parder{\L}{s}\right)\parder{}{p_i} + \L\parder{}{s} $$
on the submanifold $\W_1$, where $C^i$ are arbitrary function.

The constraint algorithm continues by demanding the tangency of $Z$ to the first constraint submanifold $\W_1$, in order to ensure that the solutions to the Lagrangian--Hamiltonian problem (the integral curves of $Z$) remain in the submanifold $\W_1$. The constraint functions defining $\W_1$ are
$$ \xi^1_j = p_j - \parder{\L}{v^j}\in\Cinfty(\W)\,. $$
Imposing the tangency condition $\Lie_Z\xi^1_j = 0$ on $\W_1$, we get
\begin{equation}\label{eq:tangency-condition}
    C^i\parderr{\L}{v^i}{v^j} = -\parderr{\L}{t}{v^j} - v^i\parderr{\L}{q^i}{v^j} - \L\parderr{\L}{s}{v^j} + \parder{\L}{q^j} + p_j\parder{\L}{s}
\end{equation}
on $\W_1$. At this point, we have to consider two different cases:
\begin{itemize}
    \item When the Lagrangian function $L$ is regular, from \eqref{eq:tangency-condition} we can determine all the coefficients $C^i$. In this case, we have a unique solution and the algorithm finishes in just one step.
    \item In case the Lagrangian $L$ is singular, equations \eqref{eq:tangency-condition} establish some relations among the functions $C^i$. In this some of them may remain undetermined and the solutions may not be unique. In addition, new constraint functions $\xi^2_j\in\Cinfty(\W_1)$ may arise. These new constraint function define a submanifold $\W_2\hookrightarrow\W_1\hookrightarrow\W$. The constraint algorithm continues by imposing that $Z$ is tangent to $\W_2$ and so on until we get a final constraint submanifold $\W_f$ (if possible) where we can find solutions to \eqref{eqn:hamiltonian_equations} tangent to $\W_f$.
\end{itemize}

Consider an integral curve $\gamma(r) = (t(r), q^i(r), v^i(r), p_i(r), s(r))$ of the vector field $Z\in\X(\W)$. We have that $A = \dot t$, $B^i = \dot q^i$, $C^i = \dot v^i$, $D_i = \dot p_i$ and $E = \dot s$. Then, equations \eqref{eq:SR-second}, \eqref{eq:SR-third}, \eqref{eq:SR-contact-one}, \eqref{eq:SR-contact-two} and \eqref{eq:SR-contact-three} lead to the local expression of \eqref{eqn:hamiltonian_equations_coordinates}. In particular,
\begin{itemize}
    \item Equation \eqref{eq:SR-contact-one} gives $v^i = \dot q^i$, namely the holonomy condition.
    \item Combining equations \eqref{eq:SR-second} and \eqref{eq:SR-contact-one}, we see that
    \begin{equation}\label{eq:dot_s}
        \dot s = \L\,,
    \end{equation}
    which corresponds to equation \eqref{eq:EL-curve-coords-2}.
    \item Equations \eqref{eq:SR-contact-three} give
    $$ \dot p_i = \parder{\L}{q^i} + p_i\parder{\L}{s} = -\left(\parder{\H}{q^i} + p_i\parder{\H}{s}\right)\,, $$
    which are the second set of Hamilton's equations \eqref{eq:Hamilton-cocontact}. These equations, on the first constraint submanifold $\W_1$, read
    $$ \frac{\d}{\d t}\left(\parder{\L}{v^i}\right) - \parder{\L}{q^i} = \parder{\L}{v^i}\parder{\L}{s}\,, $$
    which are the Herglotz--Euler--Lagrange equations \eqref{eq:EL-curve-coords-3}. Also, the first set of Hamilton's equations \eqref{eq:Hamilton-cocontact} comes from the definition of the Hamiltonian function \ref{dfn:Hamiltonian function} taking into account the holonomy condition.
    \item Combining equations \eqref{eq:SR-contact-two} and \eqref{eq:dot_s}, the tangency condition \eqref{eq:tangency-condition} yields the Herglotz--Euler--Lagrange equations \eqref{eq:EL-curve-coords-3}. It is important to point out that, if the Lagrangian function $L$ is singular, the Herglotz--Euler--Lagrange equations may be incompatible.
\end{itemize}

\section{Recovering the Lagrangian and Hamiltonian formalisms}

The aim of this section is to show the equivalence between the Skinner--Rusk formalism, presented above, and the Lagrangian and Hamiltonian formalisms.

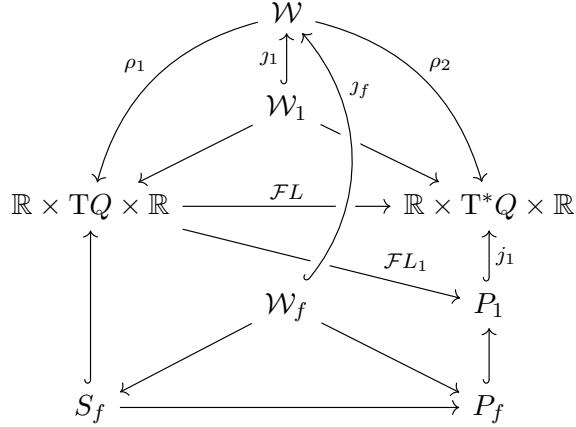
\begin{figure}[h]
 \centering
 \begin{tikzcd}
  & \W \arrow[ddr, bend left, "\rho_2"] \arrow[ddl, bend right,  swap, "\rho_1"] \\
  & \W_1 \arrow[dl] \arrow[dr] \arrow[u, hookrightarrow, "\jmath_1"] \\
  \R\times\T Q\times\R \arrow[rr, "\F L"] \arrow[rrd, pos=0.7, "\F L_1"] & & \R\times\cT Q\times\R \\
  & \W_f \arrow[dl] \arrow[dr] \arrow[uuu, bend right=40, hookrightarrow, crossing over, swap, pos=0.7, "\jmath_f"]  & P_1 \arrow[u, hookrightarrow, swap, "j_1"] \\
  S_f \arrow[rr] \arrow[uu, hookrightarrow] & & P_f \arrow[u, hookrightarrow]
 \end{tikzcd}
 \caption{Recovering the Lagrangian and Hamiltonian formalisms}
 \label{fig:diagram-contact-Skinner-Rusk}
\end{figure}

Let us denote $\jmath_1\colon\W_1 \hookrightarrow \W$ as the natural embedding. Then

$$
\rho_1\circ\jmath_1 \colon \W_1 \rightarrow \R\times\T Q\times\R
\qquad \mbox{ and } \qquad
\rho_2\circ\jmath_1\colon\W_1 \rightarrow \R\times\cT Q\times\R
$$
where
$$
(\rho_1\circ\jmath_1)(\W_1) = \R\times\T Q\times\R
\qquad \mbox{ and } \qquad
(\rho_2\circ\jmath_1)(\W_1) = P_1 \subset \R\times\cT Q\times\R\,.
$$
Also $P_1$ is a submanifold of $\R\times\cT Q\times\R$ whenever $L$ is an almost regular Lagrangian (see \cite{DeLeo2022} for a precise definition of this concept in the cocontact setting)
and we have the equality $P_1 = \R\times\cT Q\times\R$ when $L$ is regular. Since $\W_1 = \graph(\F L)$, it is diffeomorphic to $\R\times\T Q\times\R$ and $(\rho_1\circ\jmath_1)$ is a diffeomorphism. Similarly, under the assumption $L$ is almost-regular, we have
$$
(\rho_1\circ\jmath_\alpha)(\W_\alpha) = S_\alpha \subset \R\times\T Q\times\R
\qquad \mbox{ and } \qquad
(\rho_2\circ\jmath_\alpha)(\W_\alpha) = P_\alpha \subset P_1 \subset \R\times\cT Q\times\R\,.
$$
for every $\W_\alpha$ submanifold obtained from the constraint algorithm.
Let us denote by $\W_f$ the final constraint submanifold
$$ \W_f \hookrightarrow \dots \hookrightarrow \W_\alpha \hookrightarrow \dots \hookrightarrow \W_1 \hookrightarrow \W\,. $$

\begin{thm}
  Consider a path $\sigma\colon I\subset\R\rightarrow\W_1$.
  Therefore $\sigma=(\sigma_L, \sigma_H)$ where $\sigma_L = \rho_1\circ\sigma$ and $\sigma_H = \F L\circ\sigma_L$. Denote also $\sigma_0 = \rho_0\circ\sigma: I\subset\R\rightarrow\R\times Q\times\R$.
  Then:
  \begin{itemize}
    \item If $\sigma\colon I \subset\R \rightarrow\W_1$ fulfills equations \eqref{eqn:hamiltonian_equations_coordinates} on $\W_f$ then $\sigma_L$ is the prolongation of the curve $\sigma_0 = \rho_0\circ\sigma$, which is a solution of \eqref{eq:Euler-Lagrange-curve}.
      Also the path $\sigma_H = \F L\circ\widetilde\sigma_0$ is a solution to \eqref{eq:Ham-eq-cocontact-sections} on $P_f$.
    \item Conversely, let the path $\sigma_0: I\subset\R \rightarrow \R\times Q\times\R$ be a solution of \eqref{eq:Euler-Lagrange-curve} on $S_f$. Then, the section $\sigma = (\sigma_L, \sigma_H)$ which is a solution of \eqref{eqn:hamiltonian_equations_coordinates}, where $\sigma_L := \widetilde\sigma_0$ and $\sigma_H := \F L\circ\widetilde\sigma_0$. Also path $\sigma_H$ is a solution to \eqref{eq:Ham-eq-cocontact-sections} on $P_f$.
  \end{itemize}
\end{thm}

As a consequence, we obtain the following recovering theorems:

\begin{thm}
  Let $Z\in\vf(\W)$ be a solution of \eqref{eqn:hamiltonian_equations} at least on $\W_f$ and tangent to $\W_f$.
  Then:
  \begin{itemize}
    \item The vector field $X\in\vf(\R\times\T Q\times\R)$ defined by $X\circ\rho_1 = \T\rho_1\circ Z$ is a holonomic vector field (tangent to $S_f$) which is a solution to the equations \eqref{eq:Euler-Lagrange-field} with $\H = \rho_1^*E_L$.
    \item Conversely, every holonomic vector field solution of the equations \eqref{eq:Euler-Lagrange-field} can be recovered in this way from a vector field $Z\in\vf(\W)$ solution of \eqref{eqn:hamiltonian_equations} on $\W_f$ and tangent to $\W_f$.
  \end{itemize}
\end{thm}
Similarly, we also recover the Hamiltonian formalism:

\begin{thm}
  Let $Z\in\vf(\W)$ be a solution of \eqref{eqn:hamiltonian_equations} at least on $\W_f$ and tangent to $\W_f$. Then, the vector field $Y\in\vf(\R\times\cT Q\times\R)$ defined by $Y\circ\rho_2 = \T\rho_2\circ Z$ is a solution to the equations \eqref{eq:Ham-eq-cocontact-vectorfields} on $P_f$ and tangent to $P_f$.
\end{thm}

\section{Examples}

\subsection{The Duffing equation}

The Duffing equation \cite[p. 82]{Guc1983}, \cite{Wig2003}, named after G. Duffing, is a non-linear second-order differential equation which can be used to model certain damped and forced oscillators. The Duffing equation is
\begin{equation}\label{eq:duffing-eq}
    \ddot x + \delta \dot x + \alpha x + \beta x^3 = \gamma\cos\omega t\,,
\end{equation}
where $\alpha,\beta,\gamma,\delta,\omega$ are constant parameters. Notice that if $\beta = \delta \gamma = 0$, we obtain the equation of a simple harmonic oscillator. In physical terms, equation \eqref{eq:duffing-eq} models a damped forced oscillator with a stiffness different from the one obtained by Hooke's law.

It is clear that this system is not Hamiltonian nor Lagrangian  in a standard sense. However, we will see that we can provide a geometric description of it as a time-dependent contact Hamiltonian system.

Consider the configuration space $Q = \R$ with canonical coordinate $(x)$. Consider now the product bundle $\R\times\T Q\times\R$ with natural coordinates $(t, x, v, s)$ and the Lagrangian function
$$ \L\colon\R\times\T Q\times\R\longrightarrow\R $$
given by
\begin{equation}\label{eq:duffing-lag}
    \L(t, x, v, s) = \frac{1}{2}v^2 - \frac{1}{2}\alpha x^2 - \frac{1}{4}\beta x^4 - \delta s + \gamma x\cos \omega t\,.
\end{equation}

Let $\W$ be the extended Pontryagin bundle
$$ \W = \R\times\T Q\times_Q\cT Q\times\R $$
equipped with natural coordinates $(t, x, v, p, s)$. The coupling function is $\C(t, x, v, p, s) = pv$. The couple of one-forms $(\d t,\eta = \d s - p\d x)$ define a precocontact structure on $\W$. The dissipative Reeb vector field is $\Reeb_s = \tparder{}{s}$ and the time Reeb vector field is $\Reeb_t = \tparder{}{t}$. We also have that $\d\eta = \d x\wedge\d p$. The Hamiltonian function associated to the Lagrangian function \eqref{eq:duffing-lag} is the function
$$ \H = \C - \L = pv - \frac{1}{2}v^2 + \frac{1}{2}\alpha x^2 + \frac{1}{4}\beta x^4 + \delta s - \gamma x\cos \omega t\ \in\Cinfty(\W)\,. $$
We have that
$$ \d\H = \gamma\omega x\sin(\omega t)\d t + (\alpha x + \beta x^3 - \gamma\cos\omega t)\d x + (p - v)\d v + v\d p + \delta\d s\,, $$
and hence
$$ \d\H - \Reeb_s(\H)\eta - \Reeb_t(\H)\d t = (\alpha x + \beta x^3 + \delta p - \gamma\cos\omega t)\d x + (p-v)\d v + v\d p\,. $$
Given a vector field $Z\in\X(\W)$ with local expression
$$ Z = A\parder{}{t} + B\parder{}{x} + C\parder{}{v} + D\parder{}{p} + E\parder{}{s}\,, $$
equations \eqref{eqn:hamiltonian_equations} give the conditions
$$
    \begin{dcases}
        A = 1\,,\\
        B = v\,,\\
        D = -\alpha x - \beta x^3 - \delta p + \gamma\cos\omega t\,,\\
        p-v = 0\,,\\
        E = pB - \H = \L\,.
    \end{dcases}
$$
Thus, the vector field $Z$ is a \textsc{sode} and has the expression
$$ Z = \parder{}{t} + v\parder{}{x} + C\parder{}{v} + (-\alpha x - \beta x^3 - \delta p + \gamma\cos\omega t)\parder{}{p} + \L\parder{}{s}\,, $$
and we have the constraint function
$$ \xi_1 = p-v = 0\,, $$
which defines the first constraint submanifold $\W_1\hookrightarrow\W$. The constraint algorithm continues by demanding the tangency of the vector field $Z$ to $\W_1$. Hence,
$$ 0 = \Lie_Z \xi_1 = -\alpha x - \beta x^3 - \delta v + \gamma\cos\omega t - C\,, $$
determining the last coefficient of the vector field $Z$
$$ C = -\alpha x - \beta x^3 - \delta v + \gamma\cos\omega t\,, $$
and no new constraints appear. Then, we have the unique solution
$$ Z = \parder{}{t} + v\parder{}{x} + (-\alpha x - \beta x^3 - \delta v + \gamma\cos\omega t)\parder{}{v} + (-\alpha x - \beta x^3 - \delta v + \gamma\cos\omega t)\parder{}{p} + \L\parder{}{s}\,. $$
Projecting onto each factor of $\W$, namely using the projections $\rho_1\colon\W\to\R\times\T Q\times\R$ and $\rho_2\colon\W\to\R\times\cT Q\times\R$, we can recover both the Lagrangian and the Hamiltonian vector fields. In the Lagrangian formalism we obtain the holonomic vector field $X\in\X(\R\times\T Q\times\R)$ given by
$$ X = \parder{}{t} + v\parder{}{x} + (-\alpha x - \beta x^3 - \delta v + \gamma\cos\omega t)\parder{}{v} + \L\parder{}{s}\,. $$
We can see that the integral curves of $X$ satisfy the Duffing equation
$$ \ddot x + \delta \dot x + \alpha x + \beta x^3 = \gamma\cos\omega t\,. $$
On the other hand, projecting with $\rho_2$, we obtain the Hamiltonian vector field $Y\in\X(\R\times\cT Q\times\R)$ given by
$$ Y = \parder{}{t} + p\parder{}{x} + (-\alpha x - \beta x^3 - \delta p + \gamma\cos\omega t)\parder{}{p} + \left(\frac{1}{2}p^2 - \frac{1}{2}\alpha x^2 - \frac{1}{4}\beta x^4 - \delta s + \gamma x\cos \omega t\right)\parder{}{s}\,. $$
Notice that the integral curves of $Y$ also satisfy the Duffing equation \eqref{eq:duffing-eq}. Thus, we have shown that although the Duffing equation cannot be formulated as a standard Hamiltonian system, it can be described as a cocontact Lagrangian or Hamiltonian system.


\subsection{System with time-dependent mass and quadratic drag}

In this example we will consider a system of time-dependent mass with an engine providing an ascending force $F>0$ and subjected to a drag proportional to the square of the velocity.

Let $Q = \R$ with coordinate $(y)$ be the configuration manifold of our system and consider the Lagrangian function
$$ \L\colon\R\times\T Q\times\R\longrightarrow\R $$
given by
\begin{equation}\label{eq:lagrangian-quadratic-drag}
    \L(t, y, v, s) = \frac{1}{2}m(t)v^2 + \frac{m(t)g}{2\gamma}\left( e^{-2\gamma y} - 1 \right) - 2\gamma v s + \frac{1}{2\gamma}F \,,
\end{equation}
where $\gamma$ is the drag coefficient and the mass is given by the monotone decreasing function $m(t)$. Consider the extended Pontryagin bundle
$$ \W = \R\times\T Q\times_Q\cT Q\times\R $$
endowed with canonical coordinates $(t, y, v, p, s)$, the coupling function $\C(t,y,v,p,s) = pv$ and the 1-forms $\tau = \d t$ and $\eta = \d s - p\d y$. We have that $(\W,\tau,\eta)$ is a precocontact manifold, with Reeb vector fields $\Reeb_t = \tparder{}{t}$, $\Reeb_s = \tparder{}{s}$. The Hamiltonian function $\H$ associated with the Lagrangian \eqref{eq:lagrangian-quadratic-drag} is
$$ \H = \C - \L = pv - \frac{1}{2}m(t)v^2 - \frac{m(t)g}{2\gamma}\left( e^{-2\gamma y} - 1 \right) + 2\gamma v s - \frac{1}{2\gamma}F\ \in\Cinfty(\W)\,. $$
In this case, we have
\begin{multline*}
    \d\H -\Reeb_s(\H)\eta - \Reeb_t(\H)\d t =\\= v\d p + (p - m(t) v + 2\gamma s)\d v + \big(m(t) g e^{-2\gamma y} + 2\gamma v p\big)\d y + \parder{\H}{t}\d t + 2\gamma v \d s\,.
\end{multline*}
Consider a vector field $Z\in\X(\W)$ with coordinate expression
$$ Z = A\parder{}{t} + B\parder{}{y} + C\parder{}{v} + D\parder{}{p} + E\parder{}{s}\,. $$
Then, equations \eqref{eqn:hamiltonian_equations} read
$$
    \begin{dcases}
        A = 1\,,\\
        B = v\,,\\
        D = -m(t)ge^{-2\gamma y} - 2\gamma v p\,,\\
        p-m(t)v + 2\gamma s = 0\,,\\
        E = pB - \H = \L\,.
    \end{dcases}
$$
Hence, the vector field $Z$ is a \textsc{sode} and has the expression
$$ Z = \parder{}{t} + v\parder{}{y} + C\parder{}{v} + \left( -m(t)ge^{-2\gamma y} - 2\gamma v p \right)\parder{}{p} + \L\parder{}{s}\,, $$
and we obtain the constraint function
$$ \xi_1 = p-m(t)v + 2\gamma s\,, $$
defining the first constraint submanifold $\W_1\hookrightarrow\W$. Demanding the tangency of $Z$ to $\W_1$, we obtain
$$ \Lie_Z\xi_1 = -\gamma m(t) v^2 - m(t)C - \dot m(t) v - m(t) g + F\,, $$
which determines the remaining coefficient of the vector field $Z$:
$$ C = \frac{F}{m(t)} - \gamma v^2 - \frac{\dot m(t)}{m(t)}v - g $$
and no new constraints appear. Thus, we have the unique solution
$$ Z = \parder{}{t} + v\parder{}{y} + \left(\frac{F}{m(t)} - \gamma v^2 - \frac{\dot m(t)}{m(t)}v - g\right)\parder{}{v} + \left( -m(t)ge^{-2\gamma y} - 2\gamma v p \right)\parder{}{p} + \L\parder{}{s}\,. $$

We can project onto each factor of $\W$ by using the projections $\rho_1\colon\W\to\R\times\T Q\times\R$ and $\rho_2\colon\W\to\R\times\cT Q\times\R$ thus recovering the Lagrangian and Hamiltonian formalisms. In the Lagrangian formalism we get the \textsc{sode} $X\in\X(\R\times\T Q\times\R)$ given by
$$ X = \parder{}{t} + v\parder{}{y} + \left(\frac{F}{m(t)} - \gamma v^2 - \frac{\dot m(t)}{m(t)}v - g\right)\parder{}{v} + \L\parder{}{s}\,. $$
The integral curves of $X$ satisfy the second-order differential equation
$$ \begin{dcases}
    \dot x = v\,,\\
    \dot v = \frac{F}{m(t)} - \gamma v^2 - \frac{\dot m(t)}{m(t)}v - g\,,
\end{dcases} $$
which can be rewritten as
$$ \begin{dcases}
    \dot x = v\,,\\
    \frac{\d}{\d t}\left(m(t)v\right) = F - m(t)g - \gamma m(t)v^2\,.
\end{dcases} $$


\subsection{Charged particle in electric field with friction with a time-dependent constraint}

Consider a system where we have a charged particle with mass $m$ and charge $k$ in the plane immersed in a stationary electric field $\mathbf{E} = (E_1,E_2,E_3) = -\nabla\phi$ and subjected to a time-dependent constraint given by $f(t,\mathbf{q}) = 0$, where $\mathbf{q} = (x,y,z)$. Consider the phase space $\T\R^4$, endowed with natural coordinates $(x,y,z,\lambda;v_x,v_y,v_z,v_\lambda)$, and the contact Lagrangian function $\L\colon\R\times\T\R^4\times\R\to\R$ given by
\begin{equation}\label{eq:charged-particle-lagrangian}
    \L(t, \mathbf{q}, \lambda, \mathbf{v}, v_\lambda, s) = \frac{1}{2}mv^2 - k\phi(\mathbf{q}) + \lambda f(t,\mathbf{q}) - \gamma s\,,
\end{equation}
where $\mathbf{v} = (v_x,v_y,v_z)$ and $v = \sqrt{v_x^2+v_y^2 + v_z^2}$ and $\gamma$ is the friction coefficient. Since we have introduced the restriction $f(t,\mathbf{q}) = 0$ via a Lagrange multiplier, it is clear that the Lagrangian \eqref{eq:charged-particle-lagrangian} is singular.

Consider the extended Pontryagin bundle
$$ \W = \R\times\T\R^4\times_{\R^4}\cT\R^4\times\R $$
endowed with natural coordinates $(t, x,y,z, \lambda, v_x,v_y, v_z, v_\lambda, p_x,p_y,p_\lambda, s)$. We have that $(\d t,\eta)$, where
$$ \eta = \d s - p_x\d x - p_y\d y - p_z\d z - p_\lambda\d\lambda\,, $$
is a precocontact structure on $\W$. The coupling function is $\C = p_xv_x + p_yv_y + p_zv_z + p_\lambda v_\lambda$ and the Hamiltonian function $\H = \C - \L$ associated to the Lagrangian $\L$ is
$$ \H = p_xv_x + p_yv_y + p_zv_z + p_\lambda v_\lambda - \frac{1}{2}mv^2 + k\phi(\mathbf{q}) - \lambda f(t,\mathbf{q}) + \gamma s\,. $$
Thus,
\begin{align*}
    \d\H - \Reeb_s(\H)\eta - \Reeb_t(\H)\d t &= v_x\d p_x + v_y\d p_y + v_z\d p_z + v_\lambda\d p_\lambda\\
    &\quad + (p_x - mv_x)\d v_x + (p_y - mv_y)\d v_y + (p_z - mv_z)\d v_z+ p_\lambda\d v_\lambda \\
    &\quad + \left( k\parder{\phi}{x} - \lambda\parder{f}{x} + \gamma p_x \right)\d x + \left( k\parder{\phi}{y} - \lambda\parder{f}{y} + \gamma p_y \right)\d y\\
    &\quad + \left( k\parder{\phi}{z} - \lambda\parder{f}{z} + \gamma p_z \right)\d z + \big(\gamma p_\lambda - f(t,\mathbf{q})\big)\d\lambda\,.
\end{align*}
Let $Z$ be a vector field on $\W$ with local expression
\begin{align*}
    Z &= A\parder{}{t} + B_x\parder{}{x} + B_y\parder{}{y} + B_z\parder{}{z} + B_\lambda\parder{}{\lambda} + C_x\parder{}{v_x} + C_y\parder{}{v_y} + C_z\parder{}{v_z} + C_\lambda\parder{}{v_\lambda}\\
    &\quad + D_x\parder{}{p_x} + D_y\parder{}{p_y} + D_z\parder{}{p_z} + D_\lambda\parder{}{p_\lambda} + E\parder{}{s}\,,
\end{align*}
then, equations \eqref{eqn:hamiltonian_equations} yield the conditions
\begin{gather*}
    A = 1\,,\qquad B_x = v_x\,,\qquad B_y = v_y\,,\qquad B_z = v_z\,,\qquad B_\lambda = v_\lambda\,,\\
    D_x = \lambda\parder{f}{x} - k\parder{\phi}{x} - \gamma p_x\,,\qquad D_y = \lambda\parder{f}{y} - k\parder{\phi}{y} - \gamma p_y\,,\\
    D_z = \lambda\parder{f}{z} - k\parder{\phi}{z} - \gamma p_z\,,\qquad D_\lambda = f(t,\mathbf{q}) - \gamma p_\lambda\,,\\
    p_x-mv_x = 0\,,\qquad p_y-mv_y = 0\,,\qquad p_z-mv_z = 0\,,\qquad p_\lambda = 0\,,\\
    E = p_xB_x + p_yB_y + p_zB_z + p_\lambda B_\lambda - \H = \L\,.
\end{gather*}
Thus, the vector field $Z$ is
\begin{align*}
    Z &= \parder{}{t} + v_x\parder{}{x} + v_y\parder{}{y} + v_z\parder{}{z} + v_\lambda\parder{}{\lambda} + C_x\parder{}{v_x} + C_y\parder{}{v_y} +C_z\parder{}{v_z} + C_\lambda\parder{}{v_\lambda} \\ &\quad+ \left( \lambda\parder{f}{x} - k\parder{\phi}{x} - \gamma mv_x \right)\parder{}{p_x} + \left( \lambda\parder{f}{y} - k\parder{\phi}{y} - \gamma mv_y \right)\parder{}{p_y}\\
    &\quad + \left( \lambda\parder{f}{z} - k\parder{\phi}{z} - \gamma mv_z \right)\parder{}{p_z} - f(t,\mathbf{q})\parder{}{p_\lambda} + \L\parder{}{s}\,,
\end{align*}
and we get the constraints
$$ \xi_1^x \equiv p_x - m v_x = 0\ ,\qquad \xi_1^y \equiv p_y - m v_y = 0\ ,\qquad  \xi_1^z \equiv p_z - m v_z = 0\ ,\qquad \xi_1^\lambda \equiv p_\lambda = 0\,, $$
defining the first constraint submanifold $\W_1\hookrightarrow\W$. The constraint algorithm continues by demanding the tangency of the vector field $Z$ to the submanifold $\W_1$:
\begin{align*}
    \Lie_Z\xi_1^x &= \lambda\parder{f}{x} - k\parder{\phi}{x} - \gamma p_x - m C_x = 0\\
    \Lie_Z\xi_1^y &= \lambda\parder{f}{y} - k\parder{\phi}{y} - \gamma p_y - m C_y = 0\\
    \Lie_Z\xi_1^z &= \lambda\parder{f}{z} - k\parder{\phi}{z} - \gamma p_z - m C_z = 0\\
    \Lie_Z\xi_1^\lambda &= f(t,\mathbf{q}) = 0\,,
\end{align*}
thus determining the coefficients $C_x,C_y,C_z$. In addition, we have obtained the constraint function
$$ \xi_2 \equiv f(t,\mathbf{q}) = 0\,, $$
defining the second constraint submanifold $\W_2\hookrightarrow\W_1\hookrightarrow\W$.
Then, the vector field $Z$ has the form
\begin{align*}
    Z &= \parder{}{t} + v_x\parder{}{x} + v_y\parder{}{y} + v_z\parder{}{z} + v_\lambda\parder{}{\lambda} + \frac{1}{m}\left( \lambda\parder{f}{x} - k\parder{\phi}{x} - \gamma mv_x \right)\parder{}{v_x} \\
    &\quad + \frac{1}{m}\left( \lambda\parder{f}{y} - k\parder{\phi}{y} - \gamma mv_y \right)\parder{}{v_y}+ \frac{1}{m}\left( \lambda\parder{f}{z} - k\parder{\phi}{z} - \gamma mv_z \right)\parder{}{v_z} + C_\lambda\parder{}{v_\lambda} \\ &\quad + \left( \lambda\parder{f}{x} - k\parder{\phi}{x} - \gamma mv_x \right)\parder{}{p_x} + \left( \lambda\parder{f}{y} - k\parder{\phi}{y} - \gamma mv_y \right)\parder{}{p_y}\\
    &\quad + \left( \lambda\parder{f}{z} - k\parder{\phi}{z} - \gamma mv_z \right)\parder{}{p_z} + \L\parder{}{s}\,.
\end{align*}
Imposing the tangency to $\W_2$, we condition
$$ \Lie_Z\xi_2 = \parder{f}{t} + v_x\parder{f}{x} + v_y\parder{f}{y} + v_z\parder{f}{z} = 0\,, $$
which is a new constraint function $\xi_3 \equiv \dparder{f}{t} + v_x\dparder{f}{x} + v_y\dparder{f}{y} + v_z\dparder{f}{z} = \dparder{f}{t} + \mathbf{v}\cdot\nabla f = 0\,.$ This process has to be iterated until no new constraints appear, depending on the function $f$.

\begin{figure}[h]
    \centering
    \includegraphics[width=0.6\textwidth]{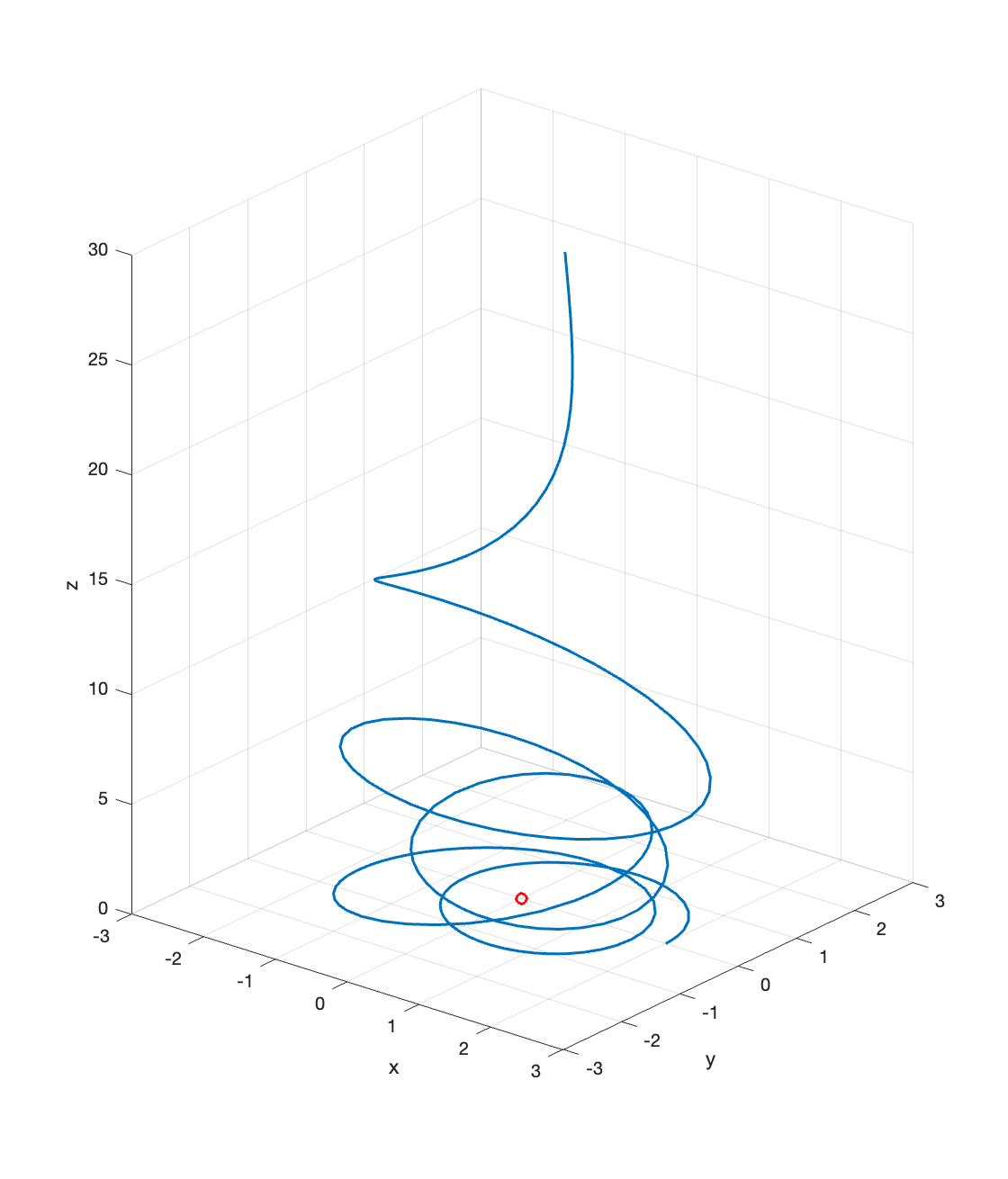}
    \caption{Trajectory of the charged particle (blue) and position of the fixed charge (red)}
    \label{fig:charged-particle}
\end{figure}

Now we are going to consider the particular case where $f(t,\mathbf{q}) = z - t$. In this case,
$$ \xi_3 \equiv v_z - 1 = 0\,. $$
Thus, imposing the tangency of $Z$ to $\xi_3$ we get the condition
$$ \Lie_Z\xi_3 = \frac{1}{m}\left( \lambda - k\parder{\phi}{z} - \gamma mv_z \right) = 0\,, $$
giving a new constraint function
$$ \xi_4 \equiv \lambda - k\parder{\phi}{z} - \gamma m = 0\,. $$
Then, the vector field $Z$ is
\begin{align*}
    Z &= \parder{}{t} + v_x\parder{}{x} + v_y\parder{}{y} + \parder{}{z} + v_\lambda\parder{}{\lambda}+ \frac{1}{m}\left( - k\parder{\phi}{x} - \gamma mv_x \right)\parder{}{v_x} + \frac{1}{m}\left( - k\parder{\phi}{y} - \gamma mv_y \right)\parder{}{v_y}\\
    &\quad + C_\lambda\parder{}{v_\lambda} + \left( - k\parder{\phi}{x} - \gamma mv_x \right)\parder{}{p_x} + \left( - k\parder{\phi}{y} - \gamma mv_y \right)\parder{}{p_y} + \L\parder{}{s}\,.
\end{align*}
Imposing the tangency condition with respect to $\xi_4$, we obtain that
$$ v_\lambda = k\left(\parderr{\phi}{x}{z}v_x + \parderr{\phi}{y}{z}v_y + \parder{^2\phi}{y^2}\right)\,. $$
The tangency condition with respect to this constraint determines the last coefficient of the vector field,
\begin{multline*}
    C_\lambda = k\Bigg( v_x^2\parder{^3\phi}{x^2\partial z} + v_y^2\parder{^3\phi}{y^2\partial z} + (1 + 2v_x)v_y\parder{^3\phi}{x\partial y\partial z} + \\ + 2v_x\parder{^3\phi}{x\partial z^2} + \parder{^3\phi}{y\partial z^2} + \parder{^3\phi}{z^3} - \left( \frac{k}{m}\parder{\phi}{x} + \gamma v_x \right)\parderr{\phi}{x}{z} - \left( \frac{k}{m}\parder{\phi}{y} + \gamma v_y \right)\parderr{\phi}{y}{z} \Bigg)\,,
\end{multline*}
and no new constraints arise. In conclusion, the vector field $Z$ has local expression
\begin{align*}
    Z &= \parder{}{t} + v_x\parder{}{x} + v_y\parder{}{y} + \parder{}{z} + k\left(\parderr{\phi}{x}{z}v_x + \parderr{\phi}{y}{z}v_y + \parder{^2\phi}{y^2}\right)\parder{}{\lambda}\\
    &\quad + \frac{1}{m}\left( - k\parder{\phi}{x} - \gamma mv_x \right)\parder{}{v_x} + \frac{1}{m}\left( - k\parder{\phi}{y} - \gamma mv_y \right)\parder{}{v_y} + C_\lambda\parder{}{v_\lambda}\\
    &\quad + \left( - k\parder{\phi}{x} - \gamma mv_x \right)\parder{}{p_x} + \left( - k\parder{\phi}{y} - \gamma mv_y \right)\parder{}{p_y} + \L\parder{}{s}\,.
\end{align*}
The integral curves of the vector field $Z$ satisfy the system of differential equations
$$ m\ddot x = -k\parder{\phi}{x} - \gamma m v_x\,,\qquad m\ddot y = -k\parder{\phi}{y} - \gamma m v_y\,,\quad z = t\,.$$

In Figure \ref{fig:charged-particle} one can see the trajectory of a charged particle with charge $k = 2\cdot 10^{-4}$ and mass $m = 1$ in the electric field induced by a charge fixed in the origin with charge $-2\cdot 10^{-4}$ and in absence of gravity. The friction coefficient is $\gamma = 0.3$ and the initial configuration of the system is $\mathbf{q}(0) = (2,0,0)$, $\mathbf{v}(0) = (0,10,0)$. As indicated above, the particle is subjected to the restriction $z = t$.

\section{Conclusions and further research}

We have generalized the Skinner--Rusk unified formalism for time-dependent contact systems. This framework allows to skip the second-order problem, since this condition is recovered in the first step of the constraint algorithm for both regular and singular Lagrangians. This makes this formalism especially interesting when working with systems described by singular Lagrangians.

The key tool of this formalism is the Pontryagin bundle $\W = \R\times\T Q\times\cT Q\times\R$ and its canonical precocontact structure. Imposing the compatibility of the dynamical equations on $\W$ we obtain a set of constraint function defining a submanifold $\W_1$, which coincides with the graph of the Legendre map, the second-order conditions and the Herglotz--Euler--Lagrange equations. We have also shown that the Skinner--Rusk formalism for cocontact systems is equivalent to both the Hamiltonian and the Lagrangian formalisms (in this last case when imposing the second order condition).

In addition, we have described in full detail three examples in order to illustrate this method: the Duffing equation, an ascending particle with time-dependent mass and quadratic drag, and a charged particle in a stationary electric field with a time-dependent constraint.

The formulation introduced in this paper will permit to extend the $k$-contact formalism for field theories with damping introduced in \cite{Gas2020,Gas2021} to non-autonomous field theories. This new formulation will permit to describe many field theories, such as damped vibrating membranes with external forces, Maxwell's equations with charges and currents, etc.

\section*{Acknowledgements}

XR acknowledges the financial support of the Ministerio de Ciencia, Innovaci\'on y Universidades (Spain), project PGC2018-098265-B-C33. 

\bibliographystyle{abbrv}
\bibliography{bibliografia.bib}

\end{document}